\documentclass[prl,aps,10pt,twocolumn,nopreprintnumbers,showpacs,tightenlines,floatfix,nofootinbib,amssymb]{revtex4-1}
\usepackage{epsfig}

\newcommand{\beqn}{\begin{eqnarray}}
\newcommand{\eeqn}{\end{eqnarray}}
\newcommand{\eq}[1]{(\ref{#1})}

\begin{document}

\title{Background magnetic field stabilizes QCD string against breaking}

\author{M. N. Chernodub}\thanks{On leave of absence from ITEP, Moscow, Russia.}
\affiliation{Laboratoire de Math\'ematiques et Physique Th\'eorique,
Universit\'e Fran\c{c}ois-Rabelais Tours,
F\'ed\'eration Denis Poisson - CNRS,
Parc de Grandmont, 37200 Tours, France\\
DMPA, University of Gent, Krijgslaan 281, S9, B-9000 Gent, Belgium}

\begin{abstract}
The confinement of quarks in hadrons occurs due to formation of QCD
string. At large separation between the quarks the QCD string breaks
into pieces due to light quark-antiquark pair creation. We argue
that there exist a critical background magnetic field $e B \simeq 16
m_\pi^2$, above which the string breaking is impossible in the
transverse directions with respect to the axis of the magnetic
field. Thus, at strong enough magnetic field a new, asymmetrically
confining phase may form. The effect -- which can potentially be
tested at LHC/ALICE experiment -- leads to abundance of $u$-quark
rich hadrons and to excess of radially excited mesons in the
noncentral heavy-ion collisions compared to the central ones.
\end{abstract}


\date{January 3, 2010}

\maketitle

{\bf Motivation.}
Noncentral heavy-ion collisions create intense magnetic fields with the magnitude of the order of the QCD scale.
According to \cite{ref:estimations} the strength of the emerging magnetic field $B$ may reach
\beqn
e B^{\mathrm{max}}_{\mathrm{RHIC}} \sim m_\pi^2\,,
\qquad
e B^{\mathrm{max}}_{\mathrm{LHC}} \sim 15 \, m_\pi^2
\label{eq:estimation}
\eeqn
at the Relativistic Heavy Ion Collider (RHIC) and at the Large Hadron Collider (LHC).
Here $e=|e|$ is the absolute value of the electron charge.

There are various potentially observable QCD effects associated with the presence of the strong magnetic field background.
One can mention a CP-odd generation of an electric current of quarks along the axis of the magnetic field (``the chiral
magnetic effect'')~\cite{ref:CME} and enhancement of the chiral symmetry breaking (``the magnetic catalysis'')~\cite{ref:magnetic:catalysis}.
The latter is related to the fact that the background magnetic field makes the
chiral condensate larger~\cite{ref:chiral:condensate}. Acting through the chiral condensate, the magnetic field also shifts
and strengthens the chiral transition~\cite{ref:chiral}.

Recently, observation of certain signatures of the chiral magnetic effect was reported by the STAR collaboration at the
RHIC experimental facility~\cite{ref:RHIC:experiment}. Some effects were also found in numerical simulations of lattice QCD.
There exists numerical evidence in favor of existence of both the chiral magnetic effect~\cite{Buividovich:2009wi} and the enhancement
of the chiral symmetry breaking~\cite{Buividovich:2008wf}. In addition, a chiral magnetization of the QCD vacuum -- discussed first
in Ref.~\cite{ref:magnetization} -- was calculated numerically~\cite{Buividovich:2009ih}. The lattice simulations have also revealed
that due to CP-odd structure of the QCD vacuum the quark's magnetic dipole moment in a strong magnetic field gets a large CP-odd piece,
the electric dipole moment~\cite{Buividovich:2009my}.

In addition to the chiral properties, the background magnetic field should also be important for confining features of QCD
despite the photons are not interacting with the gluons directly. A strong enough magnetic field affects the dynamics of the
gluons through the influence on the quarks, because the quarks are coupled to the both gauge fields~\cite{Miransky:2002rp}.
And, indeed, thermodynamic arguments of Ref.~\cite{Agasian:2008tb} suggest that the background magnetic field should
shift and weaken the confinement-deconfinement phase transition in the QCD vacuum. The confining and deconfining regions at zero
temperature are separated by a smooth crossover which is located at~\cite{Agasian:2008tb}
\beqn
e B_{\mathrm{cross}}[T=0] \sim (700\,\mbox{MeV})^2 \sim 25 \, m_\pi^2\,.
\label{eq:eBc}
\eeqn
Due to the crossover nature of the transition, the difference between confining and deconfining regions is
somewhat obscure. Therefore the confining interaction between the quarks may also be visible at the magnetic fields
that are stronger than the crossover scale~\eq{eq:eBc}.

Although in our paper we discuss QCD vacuum effects of the strong magnetic fields below the crossover scale~\eq{eq:eBc}, it is
interesting to mention that at stronger magnetic fields, $e B \gtrsim 150 \, m_\pi^2$ an exotic condensation of the $u\bar u$ pairs
should occur~\cite{Kabat:2002er}.  Effects induced by magnetic field at finite quark density are also dramatic. For example,
at $e B \gtrsim 3 m_\pi^2$ a stack of parallel $\pi^0$ domain walls should become more favorable (from an energy-related point of view)
than nuclear matter at the same density~\cite{Son:2007ny}.

\vskip 2mm
{\bf Confinement and string breaking at $\mathbf{B=0}$.}
The quark confinement exists because the chromoelectric flux (in, say, a test quark-antiquark pair)
is squeezed into a stringlike structure, the QCD string,  due to nonperturbative vacuum effects. The QCD string has a nonzero
tension $\sigma$, and therefore the energy of the long enough string is proportional to its length $R$,
\beqn
V_{\mathrm{str}}(R) = \sigma R\,,
\label{eq:Vstr}
\eeqn

\begin{figure}[!thb]
\begin{center}
\includegraphics[scale=0.48,clip=true]{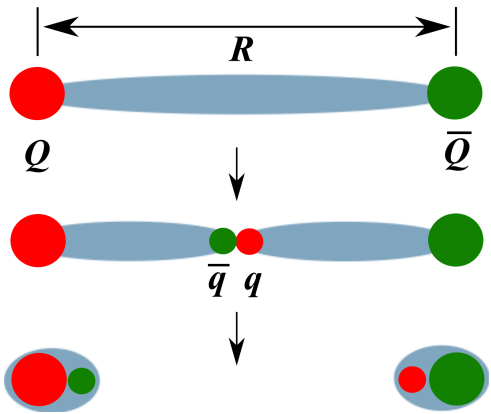}
\end{center}
\caption{The conventional string breaking: the QCD string spanned between the static
quark $Q$ and the static antiquark $\bar Q$ breaks due to light $q \bar q$ pair creation.}
\label{fig:breaking}
\end{figure}

In a quarkless QCD (i.e., in Yang-Mills theory) the energy of the QCD string~\eq{eq:Vstr}
rises infinitely with the separation between test quark and antiquark. Inevitably, the unboundedness of the potential
leads to the quark confinement. However, in the real QCD the string breaks into pieces above certain critical separation $R_{\mathrm{br}}$
because of the light quark-antiquark pair creation. The pair creation leads to the collapse of the string,
and, as a consequence, to formation of mesonic heavy-light bounds states,
\beqn
Q \bar{Q} \to Q \bar{Q} + q \bar{q} \to Q \bar{q} + q \bar{Q}\,.
\label{eq:pattern}
\eeqn
Indeed, at large separations it is more favorable to create the light pair because of the energy-related considerations:
the energy of unbroken state [given by the total mass $2 m_Q$ of the two heavy (or, test) quark sources
$Q$ and ${\bar Q}$ and the energy of the string~\eq{eq:Vstr}] is larger compared to the energy of the broken string state
(given by the mass $2 m_{Q{\bar q}}$ of the created light-heavy mesons $Q{\bar q}$ and ${\bar Q} q$).
The critical string-breaking distance $R_{\mathrm{br}}$ is thus determined by the energy balance:
\beqn
2 m_Q + \sigma R_{\mathrm{br}} = 2 m_{Q {\bar q}}
\label{eq:Vstr:br}
\eeqn
(here we neglect a weak Coulomb interaction between the quarks in the unbroken state and
we also disregard exponentially suppressed van der Waals interaction between the heavy-light
mesons in the broken state). If $R > R_{\mathrm{br}}$, then the string breaks and light-heavy mesons are
formed, Fig.~\ref{fig:breaking} (we consider here a simplest case ignoring a multiple pair production via
string fragmentation).

The string breaking was observed in lattice simulations of QCD with two flavors of equal-mass quarks~\cite{Bolder:2000un}.
An extrapolation of the lattice results to the real QCD gives the following string breaking distance~\cite{Bolder:2000un}:
\beqn
R_{\mathrm{br}} \approx 1.13 \, {\mathrm{fm}}\,,
\label{eq:Rbr}
\eeqn
where statistical and systematical errors are of the order of $0.1\, {\mathrm{fm}}$ each.

\vskip 2mm
{\bf QCD string breaking at nonzero magnetic field.}
A sufficiently strong background magnetic field should modify the dynamics of the quarks,
affecting not only the chiral features, but also influencing the confining properties of the
system. An anisotropic effect of the magnetic field on the confining scales of QCD was first
pointed out by Miransky and Shovkovy in Ref.~\cite{Miransky:2002rp}.

The energy spectrum of a {\it free} relativistic
quark in a uniform magnetic field $B$ follows a typical Landau
pattern~\cite{Johnson:1950zz}:
\beqn
\omega_{n,s_\parallel}(p_\parallel) = \pm \sqrt{p_\parallel^2 + m^2_q + (2 n + 1 - 2 s_z) |e_q| B}\,,
\label{eq:omega}
\eeqn
where $m_q$ and $e_q$ are, respectively, the mass and the electric charge of the fermion,
$p_\parallel \equiv p_z$ is the momentum of the fermion
along the direction of the magnetic field, and $s_\parallel = \pm 1/2$ is the projection of the fermion's spin
onto the axis of the magnetic field. The integer number $n=0,1,2,\dots$ labels the Landau levels. The
signs ``$\pm$'' in front of the square root in Eq.~\eq{eq:omega} refer to, respectively, particle and
antiparticle branches of the energy spectrum. The electric charges of the light $u$ and $d$ quarks are,
respectively:
\beqn
q_u = + \frac{2 e}{3}\,, \qquad
q_d = - \frac{e}{3}\,.
\label{eq:charges}
\eeqn

In a strong magnetic field the lowest Landau level (LLL) with $n=0$ and $s_z=1/2$ plays
a dominant role in the quark's motion because the excited states are too heavy. Indeed,
for the soft (low-momentum) fermions the spin flips, $s_z \to - s_z$ and jumps to the higher states
with $n \geqslant 1$ are energetically suppressed by the typical gap
\beqn
\delta E_q \sim \sqrt{2}/l_q\,,
\label{eq:gap}
\eeqn
where
\beqn
l_q(B) = 1/\sqrt{|e_q B|}\,, \qquad  q=u,d
\label{eq:aB}
\eeqn
is the magnetic length of the quark $q$.

Thus, at the LLL the motion of quarks becomes essentially (1+1) dimensional.
The quark moves along the axis of the magnetic field and the longitudinal
dynamics of a free quark is governed by one-dimensional relativistic dispersion relation:
$$\omega_{\mathrm{LLL}}(p_\parallel) = \pm \sqrt{p_\parallel^2 + m^2_q}\,.$$
The transverse dynamics (i.e., the motion of the quark in the plane orthogonal to the magnetic field) is
restricted to a region of a typical size of the order of the magnetic length~\eq{eq:aB},
\beqn
{|\delta r|}_q  \lesssim l_q(B)\,.
\label{eq:limitation}
\eeqn
In QCD the influence of the magnetic field should become significant when the magnetic
length~\eq{eq:aB} becomes comparable with a typical QCD length scale $\Lambda_{\mathrm{QCD}} \sim 1 \, {\mathrm{fm}}^{-1} \sim 200 \, {\mathrm{MeV}}$,
\beqn
l_q(B) \simeq \Lambda^{-1}_{\mathrm{QCD}} \simeq R_{\mathrm{br}}\,.
\label{eq:Bsim}
\eeqn
Here we quoted the string breaking distance $R_{\mathrm{br}}$ which was determined
by the lattice simulations~\cite{Bolder:2000un}, Eq.~\eq{eq:Rbr}.
The condition~\eq{eq:Bsim} is satisfied at
\beqn
eB_u  \simeq 6 m^2_\pi\,, \qquad eB_d = 12 m^2_\pi\,,
\label{eq:QCD:important}
\eeqn
for $u$ and $d$ quarks, respectively.

Notice that the magnitude of the magnetic field expected to be created at noncentral heavy-ion
collisions at ALICE/LHC~\eq{eq:estimation} is even greater than the thresholds~\eq{eq:QCD:important},
so that effects of the transverse quark localization may in principle be accessible in the LHC experiment.
The localization effects should be much less visible at the RHIC experiment due to the much weaker
magnetic fields~\eq{eq:estimation}.

\begin{figure}[!thb]
\begin{center}
\includegraphics[scale=0.44,clip=true]{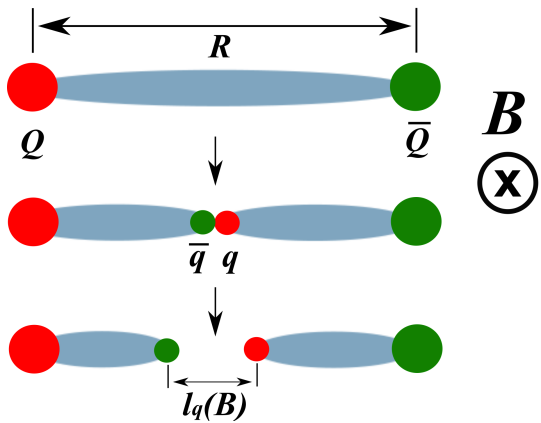}
\end{center}
\caption{The (partial) breaking of the QCD string in the background magnetic field. The distance between created
quark $q$ and antiquark $\bar q$ is restricted by the magnetic length~\eq{eq:limitation}. The axis of the magnetic field is perpendicular
to the plane.}
\label{fig:no:breaking}
\end{figure}

\vskip 2mm
{\bf What is the effect of the strong magnetic field on the process of the QCD string breaking?}
The breaking occurs due to the light $q \bar q$ pair creation. In the absence of the magnetic field
the created light quark (antiquark) gets attracted to the heavy antiquark (quark)
and the long string disappears completely, Fig.~\ref{fig:breaking}.
However, in a sufficiently strong magnetic field $B$
the distance between the light quark and antiquark
in the $\mathbf B$-transverse plane is
limited by the $B$--dependent magnetic length $l_q$, Eq.~\eq{eq:limitation}. Therefore if the string
is located in the $\mathbf B$-transverse plane, then the created $q \bar q$ pair cannot destroy the
whole string, because the light quark and antiquark
are bounded together by the magnetic field at the mutual separation $l_q$, Fig.~\ref{fig:no:breaking}.

In principle, the distance between the light quark and the light antiquark can be increased
in the transverse directions by jumping to higher Landau levels. This is, however, an energetically
costly process because of the presence of the magnetic gap~\eq{eq:gap}. Therefore the creation of the $q \bar q$ pair
from the vacuum may only remove a piece of a sufficiently long QCD string, Fig.~\eq{fig:no:breaking}.
Moreover, due to the $(1+1)$ dimensional character of the motion of the light quarks in the external magnetic
field, the $q \bar q$ pair creation leads to appearance of two nonlocal (elongated) heavy-light mesons, Fig.~\ref{fig:no:breaking}.
On the contrary, in the absence of the background magnetic field the string breaking leads to formation of the tightly-bound heavy-light mesons,
Fig.~\ref{fig:breaking}.

An increase in the magnetic field causes a decrease in the length~\eq{eq:limitation} of the string piece
that can be ``removed'' by the pair creation. Thus, as the magnetic field increases the system gets
less gain in energy due to the string breaking. Intuitively it is clear that at certain strength of the magnetic field
the pair creation become energetically unfavorable.

Let us estimate the critical magnetic field which makes the string breaking impossible.
Consider a static distantly separated quark-antiquark pair $Q \bar Q$ in the $\mathbf B$-transverse plane
(i.e., with ${\mathbf R} \equiv {\mathbf R}_\perp$ and ${\mathbf R}_\parallel = 0$).
The energy of the unbroken state is
\beqn
E_{\mathrm{string}} = 2 m_Q + \sigma(B) R\,,
\label{eq:E:string}
\eeqn
where the string tension is, generally, a function of the strength of the magnetic field $B$.

The emerging $q\bar q$ pair removes a segment of the string of the length
$l_q(B)$, and the energy of the broken state is
\beqn
E_{\mathrm{broken}} = 2 m_Q + 2 m_q(B) + \sigma(B) [R-l_q(B)]\,, \quad
\label{eq:E:broken}
\eeqn
where $m_q$ is the mass of the light quark.

The condition for the string breaking not to occur is given by the inequality
$E_{\mathrm{broken}} \geqslant E_{\mathrm{string}}$.
Using Eqs. \eq{eq:E:string} and \eq{eq:E:broken}, one gets
%the condition of the absence
%of the string breaking in the $\mathbf B$-transverse directions:
\beqn
2 m_q(B) \geqslant \sigma(B) l_q(B) \qquad \mbox{[no string breaking]}\,.
\label{eq:condition:no}
\eeqn
The equality in \eq{eq:condition:no} defines the lowest possible magnetic field (``critical no-breaking field'')
for which the string breaking does not occur.

Let us consider the condition~\eq{eq:condition:no} at zero temperature. The dynamical masses of $q=u,d$ quarks
in the QCD vacuum without a background magnetic field are
\beqn
m^{\mathrm{dyn}}_q \simeq 300\, \mathrm{MeV}\,.
\label{eq:m:dyn}
\eeqn
The external magnetic field of the practical scale~\eq{eq:estimation} makes negligible
contribution to the quarks' masses (a significant contribution to the masses
of the quarks is expected at the strength $eB \gtrsim (10\, \mathrm{TeV})^2$, Ref.~\cite{Miransky:2002rp}).
Assuming that the critical no-breaking magnetic field is lower than the crossover strength~\eq{eq:eBc},
and taking into account the fact that at the crossover all observables are smooth, we ignore an essentially
nonperturbative $B$-dependence of the string tension. We set the string tension to its phenomenological value at $B=0$,
\beqn
\sigma(B) \approx \sigma(B=0) = (440\,\mathrm{MeV})^2\,.
\label{eq:sigma:ph}
\eeqn

Using Eqs. \eq{eq:condition:no}, \eq{eq:m:dyn} and \eq{eq:sigma:ph} one finds that the QCD string fails to
break via creation of  $u \bar u$ and $d \bar d$ pairs if the magnetic fields reach the following values, respectively
\beqn
e B^{(u)}_{\mathrm{cr}} \simeq 8 m_\pi^2\,, \qquad e B^{(d)}_{\mathrm{cr}} = 16 m_\pi^2\,.
\label{eq:eB:no}
\eeqn
The critical fields~\eq{eq:eB:no} belong to the confinement region below the
$T=0$ crossover scale, Eq.~\eq{eq:eBc}, as anticipated.

At the background magnetic field $ 0 < B < B^{(u)}_{\mathrm{cr}}$ the string breaking in the $\mathbf{B}$-transverse plane
may proceed via the creation of both $u$ and $d$ quarks. In the field window $B^{(u)}_{\mathrm{cr}} < B < B^{(d)}_{\mathrm{cr}}$ the
creation of the $u\bar u$ pairs becomes energetically unfavorable, while the string breaking via the $d \bar d$ pair creation can
still occur. At $B > B^{(d)}_{\mathrm{cr}}$ the QCD string cannot be broken neither by $u \bar u$- or by $d \bar d$-pair creation.

The difference in the critical values~\eq{eq:eB:no} can be understood as follows.
According to Eqs.~\eq{eq:charges} and \eq{eq:aB},
the magnetic length $l_{d}$ of the $d$-quark is $\sqrt{2}$ times longer then the magnetic length
$l_{u}$ of the $u$-quark. Consequently, the gain in the energy due to the $d \bar d$-pair creation is
larger compared to the energy gain achieved due to emergence of the $u$-quark
pair. Thus, one needs a stronger magnetic field to suppress the $d \bar d$-pair creation compared to a field
needed to suppress the creation of a $u \bar u$-pair, $B^{(u)}_{\mathrm{cr}}  < B^{(d)}_{\mathrm{cr}}$.

\vskip 2mm
{\bf Proposal for numerical simulations.}
The effect of ``freezing'' of the QCD string breaking can be tested in the lattice simulations of QCD with dynamical fermions.
The magnetic fields of the order of~\eq{eq:eB:no} and higher were already achieved in the (quenched)
lattice simulations~\cite{Buividovich:2009wi,Buividovich:2008wf,Buividovich:2009ih,Buividovich:2009my}.
There are no principal obstacles to extend these simulations to QCD with dynamical fermions and
check the prediction \eq{eq:eB:no}.

\vskip 2mm
{\bf Possible experimental consequences.}
The freezing of the QCD string breaking in the magnetic field may also have experimental consequences. Firstly, according to the
estimation~\eq{eq:estimation} of  Ref.~\cite{ref:estimations}, the fields of the order of the critical value~\eq{eq:eB:no}
may emerge in noncentral heavy-ion collisions at the LHC. However, our considerations do not apply to the first moments of the created
quark-gluon fireball because the system stays in the deconfinement phase due to hot environment of the collision. As the plasma expands
and cools down, it hadronizes in the presence of the decaying magnetic fields. The hadronization process in the magnetic field background
involves string breaking events which are more favorable for the $u$ quarks compared to the $d$ quarks. Thus, the flavor-dependent freezing
of the string breaking in the external magnetic field leads to the $u$-quark-rich content of the hadrons created in noncentral heavy-ion
collisions compared to the central ones.

Secondly, the background magnetic field stabilizes highly excited
mesons by (i) freezing the quarks at the end-points of QCD string
and (ii) suppressing the process of the light pair creation in the
$\mathbf B$-transverse plane. Since in the noncentral collisions the
magnetic field axis is perpendicular to the reaction plane, we
expect that the magnetic field reveals itself via abundance of
radially excited mesons in the noncentral collisions compared to the
central ones. The excited mesons should dominantly be polarized in
the reaction plane, as the QCD string in such mesons should tend to
be perpendicular to the magnetic field axis. In each noncentral collision
the orientation of the reaction plane can easily be determined from
the elliptic flow of the emitted particles.

Thus, we are coming to a qualitative prediction that can in principle be checked experimentally in the heavy-ion collision experiments.
The long QCD strings should break into pieces by multiple $q\bar q$ pair creation leading to formation of generally unstable hadrons.
These primary hadrons later decay into more stable hadrons, leptons and photons, eventually forming jets~\cite{Andersson:1983ia}.
Since the particles in the jets come out essentially aligned along the original string axis, we expect that the flavor-dependent
freezing effect of the magnetic field on the QCD string breaking leads to an excess of the $u$-quark rich jets parallel to the reaction plane
of the noncentral heavy-ion collisions.

\end{document}